# A SURVEY OF FUZZY CONTROL FOR STABILIZED PLATFORMS

Said LEGHMIZI and Sheng LIU

College of Automation, Harbin Engineering University, Harbin, China
saidleg@gmail.com ; liu.sch@163.com

## ABSTRACT

*This paper focusses on the application of fuzzy control techniques (fuzzy type-1 and type-2) and their hybrid forms (Hybrid adaptive fuzzy controller and fuzzy-PID controller) in the area of stabilized platforms. It represents an attempt to cover the basic principles and concepts of fuzzy control in stabilization and position control, with an outline of a number of recent applications used in advanced control of stabilized platform. Overall, in this survey we will make some comparisons with the classical control techniques such us PID control to demonstrate the advantages and disadvantages of the application of fuzzy control techniques.*

## 1. INTRODUCTION

The popularity of the satellite receivers has significantly increased over the past decade. Today, digital programming is being delivered by a number of different companies using satellites to transmit signals to earth-based small satellite dishes. To receive the TV signal of satellites with the satellite aerial it must make an aerial of receivers point to the position of the satellite as in most instances; the consumers install the small satellite antenna dish at a fixed geographic site such as antenna at home. But if we need to install this antenna in a mobile site like cars or marine vessels, we have to use an object to maintain the antenna steady relative to a desired orientation even though the vehicle or vessel moves in different directions. This is accomplished by mounting a stabilizer platform providing rapid alignment between the satellite antenna dish targeted on the satellite and the moving vehicle.

The stabilized platform is the object which can isolate motion of the vehicle, and can measure the change of platform's motion and position incessantly, exactly hold the motorial gesture benchmark, so that it can make the equipment which is fixed on the platform aim at and track object fastly and exactly.

But the stabilized platform poses a particular problem with vessels especially in a sea and oceans. When a vessel moves in water, the ship heading may shift (yaw), the vessel may tilt along the length (pitch), or from side to side (roll). Hence the stabilizer platform must rapidly compensate for the disturbances caused by the changes in yaw, pitch and roll to maintain the small satellite antenna dish targeted on the satellite. In addition, the stabilized platform should be capable of rapid alignment so as to maintain the integrity of the received signal from the targeted satellite.

Since they began to be utilized about 100 years ago, stabilized platforms have been used on every type of moving vehicle, from satellites to submarines, and are even used on some handheld and ground-mounted devices. Its application is quite abroad and it becomes investigative hotspot in most counties all the time. With the rapid development in recent years, this technique has been widely used in the military area, such us missile guidance, airborne weapon, shipborne weapon and so on. At the same time it also been widely used in the civil area, such us public security, fire protection, surveillance, control of environment and TV reception, etc. Examples of the scientific, military, and commercial applications are shown in Figure 1. The scientific challenge in the ship carried stabilized platform design and control in cluttered environments and the lack of existing solutions was very motivating.

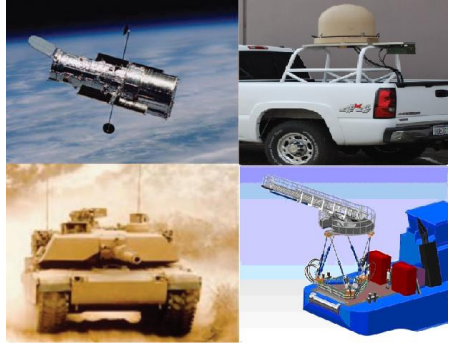

Figure 1.  Applications of stabilized platforms.

The purpose of this survey is to present the main control techniques used in the stabilized platforms and the background of some fuzzy control techniques as new paradigms and tools for solving control problems in stabilizing and control of the platform.

## 2. STABILIZED PLATFORMS RESEARCHES OVERVIEW

In recent years, the gyro-stabilized platform technology has been greatly developed by using precision machinery, micro-electronics technology [1], digital signal processing technology [2], power electronics and servo drive [3].

The foreigner top major research on the stability technology is smaller [4]. Digital and integrated system platforms have been widely used in several areas such as vehicles, ship-borne, airborne, and on-board equipment [5]. In the 40th year of the 20th century, they installed the artillery stabilizer in the tanks to reduce body vibration on the impact of the road firing. The artillery stabilizer can keep the artillery machine guns tied to stability in the required angle.

In Britain, United States and other countries advanced weapons systems, based on micro-inertial sensors for tracking, the stabilized platform has been widely used. The U.S. Navy used the quartz tuning fork gyroscope to develop the ship-based satellite communications system antenna stabilization systems [4, 6, 7].

In China Gyro stabilized platform study start late. In 80th year of the 20th century Chinese begin using the stabilized platform, until the beginning of 90th year were they started the research of development of the gyro-stabilized platform.

Although a big number of units, such as Beijing 618 laboratory, Changchun Optoelectronics department, Chinese Academy of Sciences in Chengdu, Xi'an Institute of Applied Optics, central China optoelectronics Technology Research Institute and Tsinghua University have carried out the applications in the field of research work. However, the research on the stable tracking platform technology has not achieved a breakthrough in the progress due to the higher cost of the inertia component technology. When compared with foreign countries there is still a wide gap between the research developments [4].

In our research the considered platform is a ship carried stabilized platform. The figure 2 presents the 3D presentation of the platform. To stabilize this platform we confronted to a lot of difficulties, the control difficulty and structure problems.

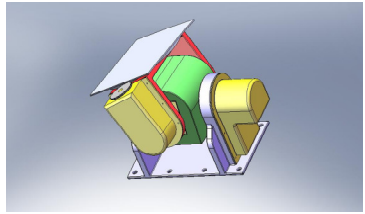

Figure 2. Stabilized platform assembly.

In our application, the complete servomechanism system comprises multiple axes. The control of such multivariable servomechanisms is, in general, not a simple problem as there exist cross-couplings, or interactions, between the different channels. In addition, this system is required to maintain stable operation even when there are changes in the system dynamics.

Also the torque disturbances in the system can be due to bearing and motor friction, unbalanced aerodynamics, vibration forces from onboard mechanisms, and spring torque forces from wires or flexures. In the stabilized platform systems, the basic requirement is to have very good disturbance rejection capability, which calls for very high gain feedback loops but at the same time the bandwidth cannot be extended too much because other considerations such as resonance in the mechanical structure, noise, and modulating frequencies. Presence of inherent nonlinearities such as stiction friction, saturation of actuators, etc. is also must be taken into account. So we will try to apply the intelligent controller in order to deal with the system dynamics uncertainty and obtain the best performance of the system.

## 3. FUZZY CONTROLLER FOR STABILIZED PLATFORM

Surveys found in [3, 4, 6, 7, 8, 9, 10, 11, 12, 13, 14, 15] provide an extensive information on various control techniques used for stabilized platform. Mostly conventional methods of control synthesis have been reported for such applications. For conventional design methods, well-known techniques such as Bode plots are applied, Modern synthesis tools such as linear quadratic regulator (LQR) or linear quadratic Gaussian with loop transfer recovery (LQG/LTR) and H∞ control methodologies have also been used in some applications [10, 11]. These methods couldn't obtain the desired performance required by the platform stabilization, due to the big disturbance in the vehicle and the nonlinearity of the platform system. For achieving this control demands and application in practice, other methods of control or combine many methods together were applied. In this work we focus on intelligent control system such us fuzzy control system [16, 17] to attempt the best control characteristics.

Intelligent control is probably the wave of the future. Its simplicity and power are the main features that make this control very fascinating. The key test, however, is when this fascination is transferred into real applications. However, a promising merge between the classical control with limited performance and the intelligent control is probably the solution that can be acceptable by the control engineers. The papers [8], [18] and [19] provided an overview of various different combinations of the classical controllers and the intelligent controllers. The intelligent control is usually active all the time and is in command. But if the system performance is judged to be unacceptable from the stability point of view, the intelligent control is turned off and the classical control is turned on [20].

### 3.1 TYPE-1 FUZZY CONTROLLER

Over the past two decades, the field of fuzzy controller applications has broadened to include many industrial control applications, and significant research work has supported the development of fuzzy controllers. Type-1 fuzzy logic systems (FLSs), constructed from type-1 fuzzy sets introduced by Zadeh in 1965, have been successfully applied to many fields. In 1974, Mamdani [21] pioneered the investigation of the feasibility of using compositional rule of inference that has been proposed by Zadeh [22], for controlling a dynamic plant. A year later, Mamdani and Assilian [23] developed the first fuzzy logic controller (FLC), and it successfully implemented to control a laboratory steam engine plant.

C.C Lee [16, 17] presented the main ideas underlying the Fuzzy Logic Controller. To highlight the issues involved, Figure 3 shows the basic configuration of an FLC, which comprises four principal components: a fuzzification interface, a Knowledge base, decision making logic, and a defuzzification interface.

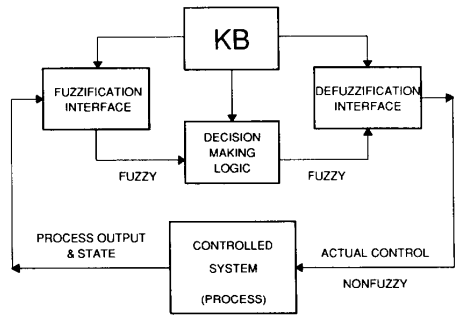

Figure 3. Basic configuration of fuzzy logic controller (FLC).

There are two major types of Type-1 fuzzy systems: Mamdani fuzzy systems and Takagi-Sugeno (TS) fuzzy systems. The main difference between the two lies in the consequent of fuzzy rules. Mamdani fuzzy systems use fuzzy sets as rule consequent, in [24], the authors provide detailed approximation accuracy analysis for some Mamdani fuzzy systems. Whereas, TS fuzzy systems employ linear functions of input variables as rule consequent, in [25], the author provides detailed calculation of TS systems and he pointed out the importance of these two types of fuzzy systems in control and modelling.

In Stabilized platforms control the author in paper [9] introduces a Fuzzy-knowledge-based controller (FKBC) design, which presents a good methodology to overcome above difficulties. Then he implemented a fuzzy control system to control inertial rate of LOS stabilized platform. Finally, he presented the design and implementation aspects of this particular FKBC and the performance achieved with respect to the performance achieved using a conventional controller.

However, research has shown that the ability of type-1 fuzzy sets to model and minimize the effect of uncertainties is limited. A reason may be that a type-1 fuzzy set is certain in the sense that for each input, there is a crisp membership grade.

### 3.2 TYPE-2 FUZZY CONTROLLER

Zadeh [26] introduced the concept of type-2 fuzzy sets as an extension of the concept of an ordinary fuzzy set, i.e., a type-1 fuzzy set. Type-2 fuzzy sets have grades of membership that are themselves fuzzy. A type-2 fuzzy set is characterised by a concept called footprint of uncertainty (FOU). Consequently, the membership grade of each element in a type-2 fuzzy set is a fuzzy set in [0, 1], unlike a type-1 set where the membership grade is a crisp number in [0, 1].

In [27] the authors introduce the concept of Type-2 fuzzy sets and compare the Type-2 fuzzy set preliminaries with the ordinary Type-1 fuzzy set. Two theorems were proved: the first one helps to generate the general form of the Type-2 operations, the second provides an approach to transfer the Type-1 fuzzy uncertainties to the Type-2 system.

The authors in [28] gave a presentation the schematic diagram of a type-2 FLS shown in figure 4. A type-reducer is needed to convert the type-2 fuzzy output sets into type-1 sets before they are processed by the defuzzifier to give a crisp output. Since type-2 FLSs provide an extra mathematical dimension compared with type-1 FLSs, they are very useful in circumstances where it is difficult to determine an exact membership grade for a fuzzy set. Hence, they can be used to handle more system uncertainties and have the potential to outperform their type-1 counterparts. To date, type-2 FLSs have been used successfully in decision making [29], control of mobile robots [30], pre-processing of data [31], noise cancellation [32], time-series forecasting [33], survey processing [34], nonlinear system identification [35] and control [36, 37], etc.

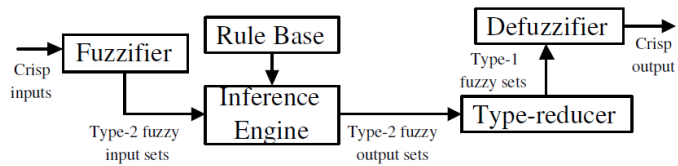

Figure 4. A type-2 fuzzy logic system.

H.X. Li et al. [38-45], also propose a concept of fuzzy controllers called probabilistic fuzzy logic system (PFLS) or 3D fuzzy logic system for the modelling and control problems. Similar to the ordinary fuzzy logic system (FLS), the PFLS consists of the fuzzification, inference engine and defuzzification operation to process the fuzzy information. Different to the FLS, it uses the probabilistic modelling method to improve the stochastic modelling capability. By using a three-dimensional membership function (MF), the PFLS is able to handle the effect of random noise and stochastic uncertainties existing in the process.

This kind of controller has not been applied in any application on stabilized platforms. Thus, in our future research we will apply the Type-2 FLS controller for stabilizing the platform.

### 3.3 COMBINATION OF THE FUZZY CONTROLLER WITH OTHERS CONTROLLERS

There are several types of control systems that use FLC as an essential system component. To increase the performances of the fuzzy logic controllers, fuzzy logic adaptation has been explored. So many combinations of the fuzzy logic and others controllers were proposed. The majority of applications during the past two decades belong to the class of fuzzy PID controllers. These fuzzy controllers can be further classified into three types: the direct action (DA) type, the gain scheduling (GS) type and a combination of DA and GS types. The majority of fuzzy PID applications belong to the DA type; here the fuzzy PID controller is placed within the feedback control loop, and computes the PID actions through fuzzy inference. In GS type controllers, fuzzy inference is used to compute the individual PID gains and the inference is either error driven self-tuning [46] or performance-based supervisory tuning [47]. In addition to the common Mamdani-type PI structure, several other structures using one- or three-input controllers have been reported.

In [48], the authors applied the Hybrid adaptive fuzzy controller. They realize a DSP based Real-Time Implementation of this method for the Tracking Control of Servo-Motor Drives and they prove the performance ameliorations of this method. The hybrid adaptive fuzzy controller employs a fuzzy logic controller integrating an H∞ tracking controller. The H∞ tracking controller is capable of taking care of the robust stabilization and disturbance rejection problems, his objective is to obtain high-performance servo tracking; therefore, the control problem is to force the actual output to follow a predetermined bounded reference trajectory. The tracking performance indicates how closely the servo motor follows a desired position trajectory, velocity trajectory or acceleration trajectory. The fuzzy logic controller is used as principle components of the adaptive fuzzy controller.

In the paper [49], the author investigates different fuzzy PID controller structures, including the Mamdani-type controller. By expressing the fuzzy rules in different forms, each PID structure is distinctly identified. Then, he gives different fuzzy PID structural elements as presented in the figure 5.

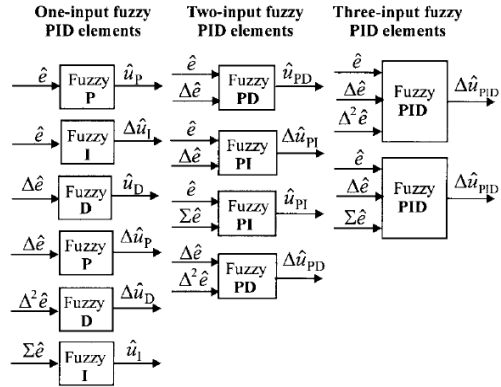

Figure 5. Fuzzy PID structural elements.

In [8], the authors apply the fuzzy PID controller for LOS Stabilized System and demonstrated using the experiment that when the system operates under the linear scope of the process, we could obtain the desired performance requirement only by the conventional PID control. But under the conditions with nonlinear constraints and uncertainties only using the conventional PID control is very difficult to realize the system high accuracy and movement stationarity. In order to satisfy the requirement for real-time, high stabilization precision and applicability, the hybrid adaptive fuzzy PID control law is introduced to realize the non-error adjustment. Figure 6 shows the structure of hybrid AFPID control proposed by the authors.

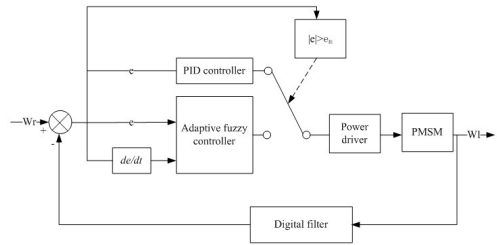

Figure 6. Structure of hybrid adaptive fuzzy PID controller.

## CONCLUSION

This paper has considered the problem of controlling a stabilized platform. We have presented an overview about the stabilized platforms researches and about the different types of fuzzy controllers applied in the stabilized platforms.

In our work we confronted to a lot of difficulties, the control difficulty and structure problems. Intelligent control is the essential tool employed in the control design, and demonstrated to be an effective solution to improve the command tracking response, the disturbance rejection response and the system robustness. Fuzzy controller is one of the most applied intelligent controllers. This is due to his simplicity and robustness.

Based on this literature review, we can argue that different fuzzy controller structures are possible in the context of knowledge representation, and that they should be evaluated with respect to their functional behaviors. Therefore, in this paper we gave an evaluation of different types of fuzzy controllers, including the combination of this controller with conventional controllers.

Our future work will focus on the design of a types-2 fuzzy control which can be used to control the proposed ship carried stabilized platform.

## AUTHORS PROFILE

**Sheng LIU**, dean of the automation college in Harbin Engineering University, director of automation society in China, his interests are stochastic process  control, the theory and application of robust control system, electromagnetic compatibility, digital signal process, optimal estimation and control of stochastic system.

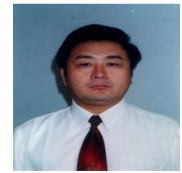

**Said LEGHMIZI** was born in Algeria in 1983. He received his Engineering diplomat degree in electrical engineering from the military polytechnic school of Algeria in 2007. He then joined Harbin engineering university as a research staff while working toward a Ph.D. degree. His research interests include   the control engineering and intelligent control systems.

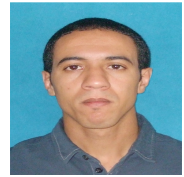